\documentclass{PoS}[20pt]
\usepackage[export]{adjustbox}
\usepackage[comma,numbers,longnamesfirst]{natbib}
\usepackage{placeins}
\usepackage{subfigure}
\usepackage{sidecap}
\usepackage{graphicx}
\usepackage{lineno}

\title{Probing Cosmic-ray Propagation with TeV Gamma Rays from the Sun Using the HAWC Observatory}

\ShortTitle{HAWC ICRC 2017 Contributions}

\author{\speaker{Mehr Un Nisa}\\
 For the HAWC Collaboration\thanks{Complete list of authors at http://www.hawc-observatory.org/collaboration/icrc2017.php}\\
        University of Rochester\\
        E-mail: \email{mnisa@ur.rochester.edu}}

\abstract{Cosmic rays in the inner solar system are subject to deflection by both the geomagnetic and interplanetary magnetic fields, and simultaneously interact with the Sun's photosphere resulting in the production of gamma rays. This phenomenon can be studied by observing the deficit ("shadow") in the cosmic ray flux from the direction of the Sun and searching for an excess photon signal above the isotropic background. The High Altitude Water Cherenkov (HAWC) Observatory in Mexico has been taking data on the solar disk at TeV energies since the end of 2014. We present our first efforts to estimate the luminosity of TeV gamma rays from the Sun which can be used to place limits on the production mechanisms, including astrophysical processes and beyond the standard model predictions.}

\FullConference{35th International Cosmic Ray Conference ICRC2017\\
		10-20 July, 2017\\
		Bexco, Busan, Korea}

\begin{document}

\section{Introduction}
Cosmic ray particles have been known to interact with the solar atmosphere, resulting in a time-dependent flux of gamma rays from the quiescent Sun. First detected by EGRET \cite{2008A&A...480..847O} and later by Fermi-LAT \cite{0004-637X-734-2-116}, the morphology of the flux reveals two distinct components:\\
$(i)$ A photon halo extending up to $20^{\circ}$ around the Sun\cite{0004-637X-734-2-116}. It is a result of cosmic ray electrons undergoing Inverse Compton (IC) scattering in the heliosphere. The intensity of gamma rays in the halo decreases as $~1/\theta$ from the center of the Sun. Fermi-LAT measurements have resolved the IC halo below $10$ GeV.\\
$(ii)$ Emission from the solar disk with an angular size of roughly $0.5^{\circ}$ in diameter. The origin of gamma rays from the solar disk is high energy cosmic rays colliding with protons in the solar atmosphere. These interactions produce neutral pions which decay into gamma rays. Charged pions are also produced and subsequently decay into muons and neutrinos. In addition, photo-disintegration of heavy cosmic ray nuclei in the Sun's photosphere is also considered to produce high energy gamma rays from the Sun \cite{2017PhRvD..95f3014B}

A distinctive feature of cosmic ray observations from the Sun's direction is the deficit in the flux due to the Sun absorbing the incoming particles. The deficit or Sun shadow overlaps with the solar disk and also varies with energy, making it difficult to find any excess of gamma rays from the Sun. This work introduces how HAWC disentangles the shadow from the solar disk and searches for an excess of gamma rays from the Sun's direction.

\subsection{Previous detections and TeV prospects} 
\begin{SCfigure}
\centering
\includegraphics[width=0.55\textwidth]{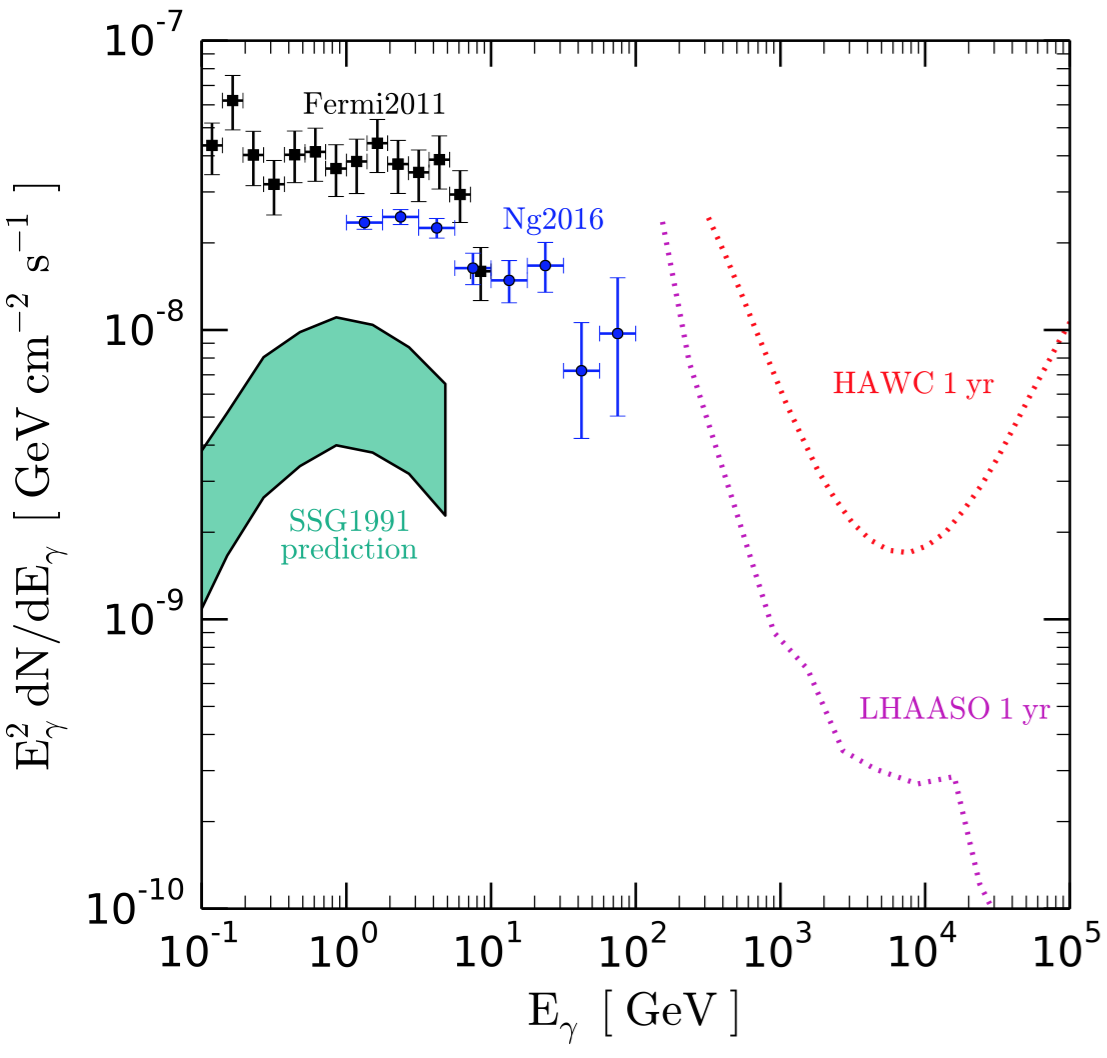}
\caption{Solar $\gamma$-ray observations with Fermi-LAT data: \textit{Black points} are the first results; \textit{blue points} are a follow up analysis of Fermi-LAT data by Ng et al \cite{Ng:2015gya} \textit{Green band}: Theoretical prediction by Seckel et al. \textit{Dotted:} TeV prospects for HAWC and LHAASO. Image courtesy: Zhou et al. \cite{2016arXiv161202420Z}}
\label{gamma}
\end{SCfigure}

The spectrum of gamma rays from the Sun has been measured up to $100$ GeV and comparisons with theoretical predictions show that the emission mechanisms are not well understood \cite{Ng:2015gya}. Fig. \ref{gamma} shows the GeV flux deduced from six years of Fermi-LAT data and a theoretical band predicted by Seckel, Stanev and Gaisser \cite{1991ApJ...382..652S}. Despite allowing for flux enhancement by magnetic fields, the prediction falls short of observations. Follow-up observations at higher energies could aid in studying the disagreement further and unravel the complete physical mechanism behind solar gamma-rays. As mentioned earlier, solar magnetic fields can enhance the flux of gamma rays from the disk \cite{2016arXiv161202420Z}, however, the magnitude of enhancement is uncertain. Observation of gamma rays at very high energies can directly measure this enhancement.

Since, gamma rays follow the spectrum of the underlying cosmic rays, extending the measurement into the TeV range provides an additional tool to probe cosmic rays in the solar neighborhood where direct measurements are not possible. Banik et al. \cite{2017PhRvD..95f3014B} also show that the TeV gamma ray spectrum is sensitive to the mass composition of cosmic rays near the knee. This lays out the possibility of resolving the uncertainties in the spectrum of ultra high energy cosmic rays by measuring solar gamma rays. Moreover, the hadronic gamma rays are accompanied by neutrinos of almost the same flux\cite{2017PhRvD..95f3014B}, offering prospects for multi messenger astronomy in the inner solar system. 

As seen in Fig. 1, a simple extrapolation of GeV data shows that a spectrum extending into the TeV range maybe within the reach of HAWC's sensitivity. A back of the envelope calculation, under very optimistic assumptions, showed that HAWC might detect gamma rays from the Sun in six years. While HAWC's collection area is limited, it is in a unique position to pursue the measurement of solar gamma rays because of its design. Cherenkov Imaging Telescopes cannot view the Sun whereas a water Cherenkov observatory has no such constraints.     

\section{High Altitude Water Cherenkov (HAWC) Observatory}
\subsection{The Detector}
The HAWC observatory is a wide field of view array of 300 water Cherenkov detectors (WCDs) covering an area of 22,000 m$^2$. It is located 4100 m above sea-level at $19^{\circ}$N near Volcano Sierra Negra, Mexico. The observatory has an instantaneous field of view of 2 sr and is designed to be sensitive to extensive air showers of secondary particles which are created by incoming cosmic particles interacting in the atmosphere. The high altitude enables HAWC to detect cosmic rays and gamma rays with energies from 300 GeV to more than 100 TeV. 
Each WCD consists of a large steel tank 4.5 m high and 7.3 m in diameter. The tank is lined with a plastic bladder filled with purified water and four photo-multiplier tubes (PMTs) attached to the bottom. As the cascade of particles from an air shower passes through the water tanks, it produces Cherenkov light which triggers the PMTs inside each detector.
\begin{figure}[h]
\begin{minipage}{0.55\textwidth}
\centering
  \includegraphics[width=0.75\textwidth]{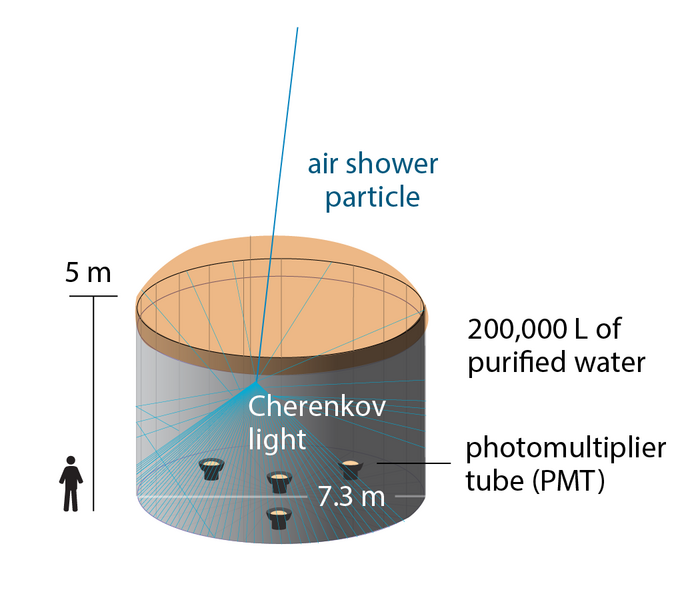}
  
\end{minipage}
\begin{minipage}{0.55\textwidth}
\centering
  \includegraphics[width=0.99\textwidth]{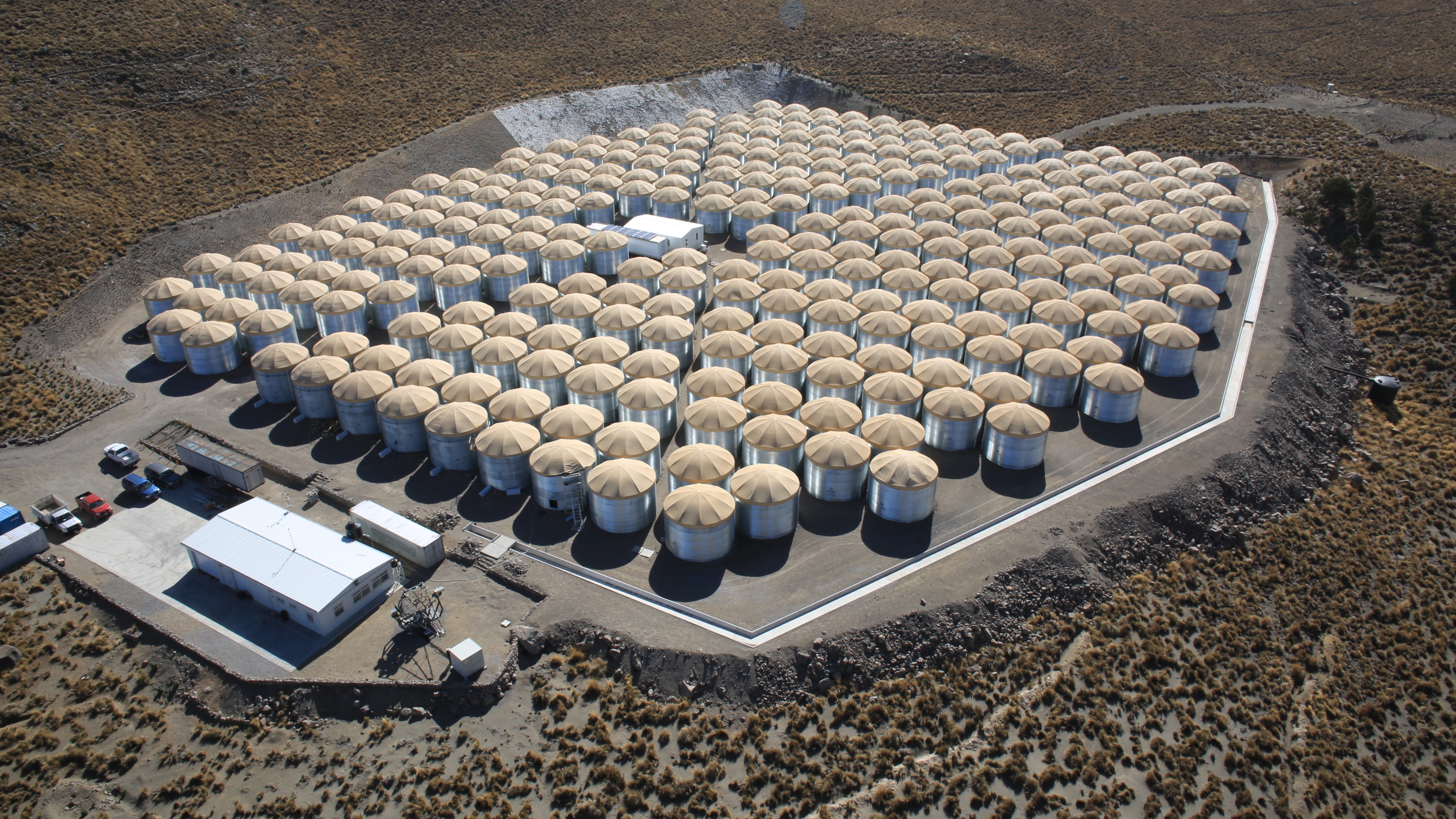}
  \end{minipage}
\caption{\textit{Left}:Schematic of the Water Cherenkov Detector with PMT's inside. \textit{Right}:The HAWC array.}
\end{figure}  

The individual PMTs have a trigger rate of $30 - 100$ kHz. The signals from each PMT are amplified, time-ordered, and their exact effective charge determined before a candidate event is processed further. After removing any incorrectly calibrated hits, the data are reconstructed to obtain the location of the shower core, the shower direction, primary particle energy and primary particle type (gamma ray or a hadron). For details of the reconstruction procedure see Ref. \cite{2017arXiv170101778A}. 

\subsection{Gamma-Hadron Separation}
In order to differentiate cosmic ray showers from gamma ray showers we employ several topological cuts making use of the fact that hadronic showers display more ``clumpy" and spatially separated charge distributions whereas the gamma showers deposit charge in a more compact and evenly spread distribution around the core. An example of a topological discriminator is \textit{Compactness}, defined as the ratio of the number of hits produced by a shower to the largest effective charge measured in a PMT outside a radius of $40$ m from the core.  This quantity is typically small for cosmic ray events because of a higher probability of a large effective charge more than $40$ m away from the core. The cuts are optimized for each reconstruction phase of the data and their efficacy varies with shower size and energy.

%
%

\section{Observation of the Sun}
To search for potential gamma rays from a given point in HAWC's field of view, we produce sky-maps of the data, quantifying the excess or deficit of CR and gamma ray counts pixel by pixel with respect to an isotropic background expectation.  To assign an estimated energy to a given set of events, we split the data into bins according to the fractional number of PMT's hit in each shower. A large number of hits corresponds to a large shower size and therefore higher energy. The data are divided into nine energy-proxy bins that allow us to analyze the flux as a function of primary energy. Using the HEALPix library \cite{2005ApJ...622..759G}, the sky is pixelized into an equal-area grid in equatorial coordinates. In order to discern features larger than the pixel size, we apply an angular smoothing that takes the event counts in each pixel and adds the counts from all pixels within a radius $\theta$. To produce the Sun maps, we use a smoothing angle of $1^\circ$ and subtract the calculated coordinates of the Sun from the coordinates of each event so that the final map is centered on the equatorial position of the Sun. 

The excess or deficit is calculated by comparing the data map at each sky position, given by right ascension and declination, to a reference map of isotropic counts. This gives the relative intensity as,
\begin{equation}
\label{eq3}
\delta I = \frac{N(\alpha_i,\delta_i) - \langle N(\alpha_i,\delta_i)\rangle}{\langle N(\alpha_i,\delta_i)\rangle}
\end{equation}
where $N(\alpha,\delta)$ is the data map and $\langle N(\alpha,\delta)\rangle$ is the isotropic reference map. The significance $\sigma$ describes the deviation of the data from the expectation in each bin of the isotropic map and is calculated using the techniques described in Refs.  \cite{1983ApJ...272..317L,2003ApJ...595..803A}.

\subsection{Sun shadow}
\begin{figure}[h!]
   \begin{minipage}{0.5\textwidth}
   \centering
    \includegraphics[width=0.85\textwidth]{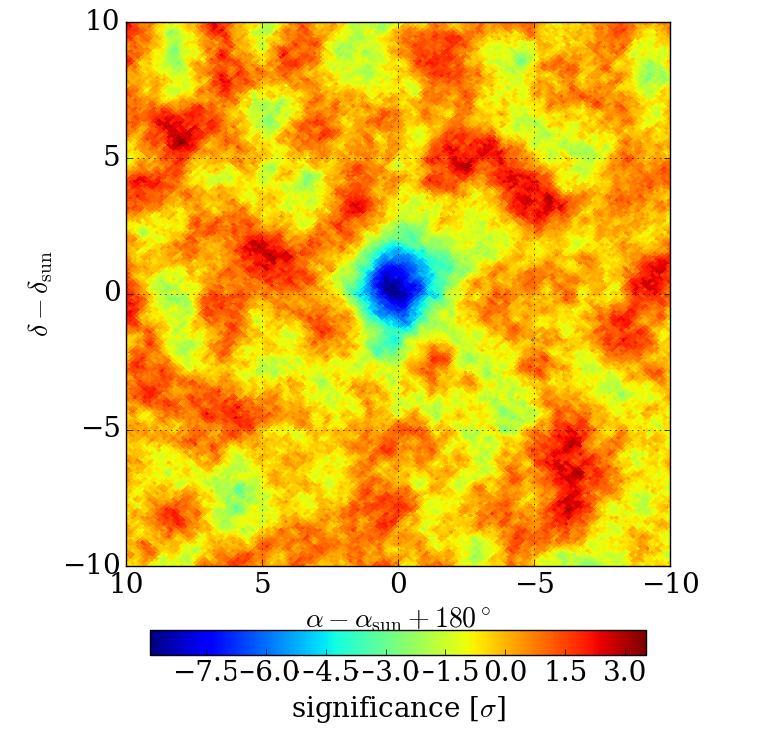}
    \end{minipage}
    \begin{minipage}{0.5\textwidth}
    \centering
    \includegraphics[width=0.85\textwidth]{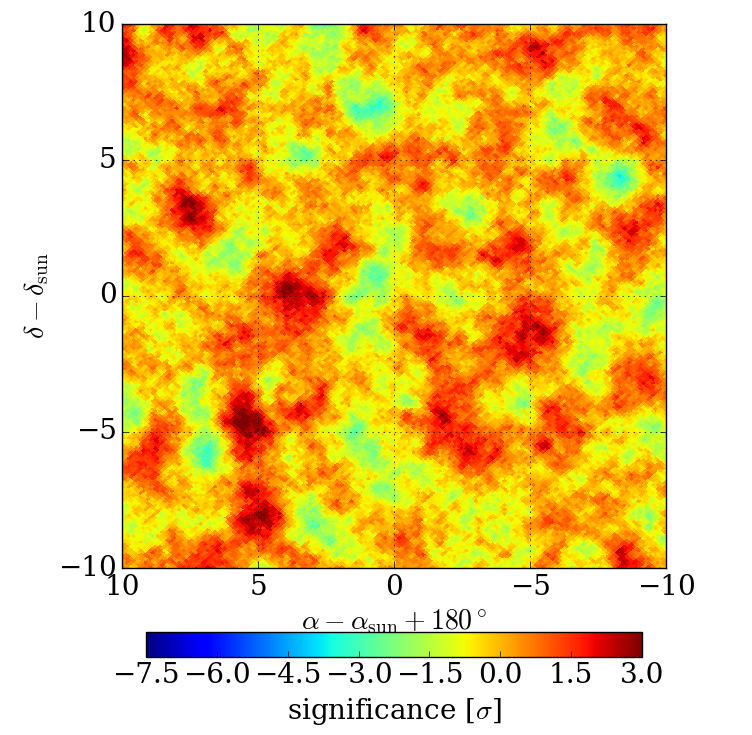}
    \end{minipage}
    \caption{Significance map of the region around the Sun in a Sun-centered (equatorial) coordinate system where the equatorial coordinates of the Sun are subtracted from the equatorial coordinates of each event.  The left panel shows all data, the right panel shows the data after applying a gamma/hadron separation cuts that reject cosmic rays}.
   \label{fig:shadow}
\end{figure}

The Sun shadow is a deficit in events with respect to the isotropic expectation and is the result of the Sun blocking out part of the incoming flux of cosmic rays as observed from a ground-based detector. This can be seen in Fig. \ref{fig:shadow}, which shows a map of bin 5 (median energy ~$20$ TeV) of over $10^{9}$ cosmic-ray events  in Sun-centered coordinates produced from 17 months of data collected by HAWC between November 2014 and June 2016. On the other hand, when the same map is made after applying gamma-hadron cuts - which remove a significant fraction of cosmic rays  leaving behind $~10^7$ events - the shadow disappears (Fig. \ref{fig:shadow}). Ideally, gamma rays from the solar limb and halo should appear as a ring and a Gaussian respectively, centered on the Sun. In the absence of any such significant feature in the data, we go on to place preliminary upper limits on gamma rays from the Sun as described in the subsequent sections.
\FloatBarrier

\subsection{Search for gamma-ray excess around the Sun}
We compare the counts in maps with gamma-hadron cuts and without the cuts to get a total excess of gamma rays above background in the data. We can write the relative intensities for the two maps as sums of their individual gamma and CR components.
\begin{equation}
\label{RI}
\delta I_{CR} = \gamma + CR
\end{equation}

For the map with gamma-hadron separation, the number of photons and CR are reduced according to the efficiency of the gamma-hadron cuts. The efficiencies $\epsilon_{\gamma}$ and $\epsilon_{CR}$ define the respective fractions of photons and cosmic rays retained after the gamma-hadron cuts are applied. These numbers from each bin are obtained from simulation \cite{ 2017arXiv170101778A}

\begin{equation}
\label{gh}
\delta I_{\gamma} = \epsilon_{\gamma}\gamma + \epsilon_{CR}CR
\end{equation}

Using equation \ref{RI} and \ref{gh}, we calculate the observed number of excess gamma counts in each bin. The relative intensities $\delta I_{CR/\gamma}$ can be read off from the respective maps and the $\epsilon_{CR/\gamma}$ are known. Solving for $\gamma$, we get

\begin{equation}
\label{gam}
\gamma = \frac{\delta I_{\gamma} - \epsilon_{CR}\delta I_{CR}}{\epsilon_{\gamma} - \epsilon_{CR}}
\end{equation}

\begin{SCfigure}
\centering
\includegraphics[width=0.6\textwidth]{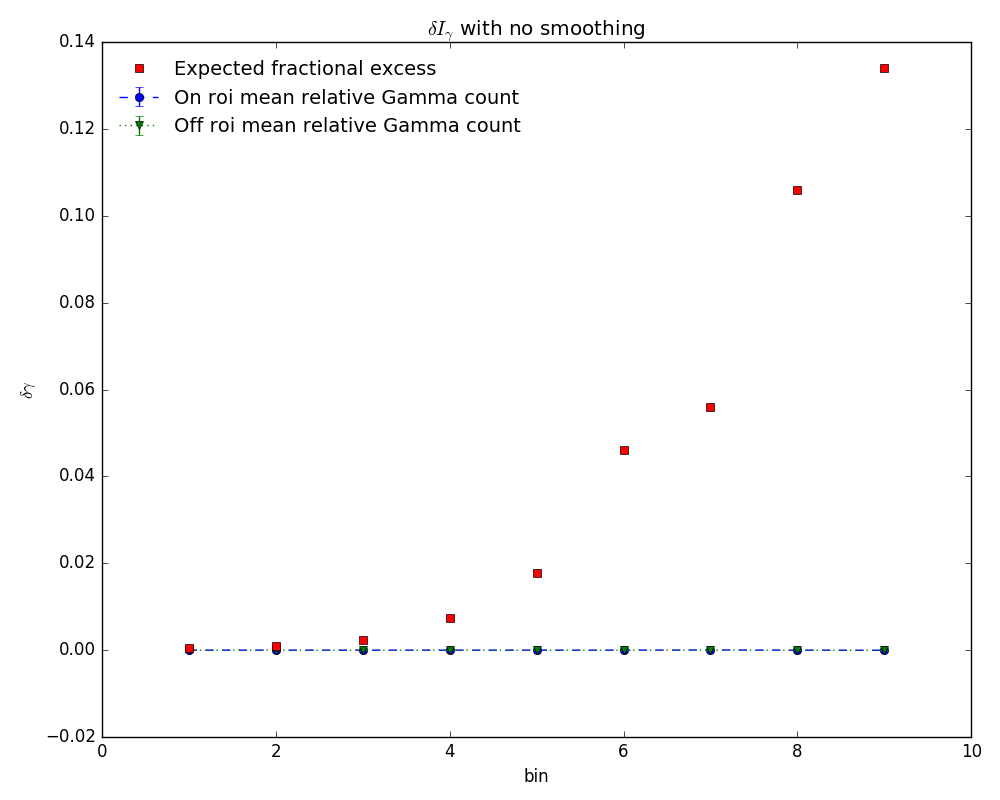}
\caption{Gamma ray excess as a function of analysis bin. On ROI represents a $5^{\circ}$ region of interest around the Sun. Off ROI is a region of similar size not centered at the Sun. The red squares show the expected excess from a Sun-like source as obtained from simulation. No significant excess is observed in the data.}
\label{fig:excess}
\end{SCfigure}

Figure \ref{fig:excess} shows the excess points for each bin. The maximum fluctuation above the background seen any bin is well below ~$1\%$ and not significant in comparison to the expected magnitude of the excess. By converting these counts into a flux and relating it with a simulated flux from a Sun-like source at different declinations, we can place upper limits on gamma rays from the Sun.

We calculate the flux that the detector would see from a point source following a simple power law. The index and the normalization of the power law, $-2.3$ and $4.36$ x $10^{-12}$/TeVcm$^2$s, respectively,  were chosen so as to be consistent with the previous measurements of solar gamma rays by Fermi-LAT\cite{0004-637X-734-2-116,Ng:2015gya}. The expected flux and the observed flux in each energy bin are related by the ratios of the observed counts and expected counts, weighted by a factor that accounts for the uncertainty in energy in each bin,
\begin{equation}
w_i = \frac{E_i}{\langle N_i\rangle}\\
\end{equation}
where $E_i$ are the expected number of events in bin $i$ (obtained from simulated source) and $\langle N_i\rangle$ is the number of background events in bin $i$.
\subsection{First Upper Limits on TeV gamma rays from the Sun}
\begin{figure}[h]
\centering
\includegraphics[width=0.92\textwidth]{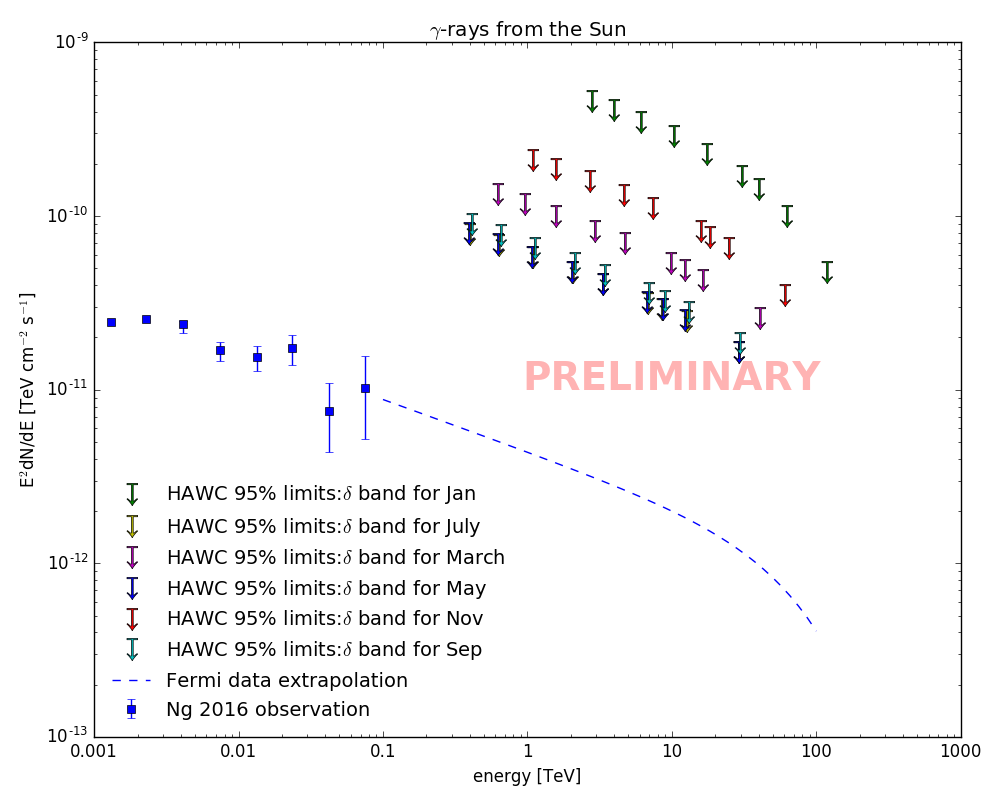}
\caption{Upper limits on differential flux of gamma ray emission from the Sun for six different declination bands to account for Sun's varying position during the year. The observed differential flux in Fermi data by Ng et al. \cite{Ng:2015gya} are also shown along with an extrapolation between $100$ GeV and $100$ TeV}
\label{yo}
\end{figure}

We obtain an estimate of the observed flux and use the errors to place $95\%$ confidence level upper limits as shown in Fig. \ref {yo}. The envelope of limits indicates that the upper limits are dependent on the declination of the Sun which varies with time. Placing an absolute, declination-independent limit requires different analysis tools and will be the subject of future work.

\FloatBarrier

\section{Conclusion}
Cosmic ray interactions around the Sun have been proposed as a source of TeV gamma rays. We performed our first search for solar gamma photons in the TeV regime in $507$ days of HAWC data. No significant excess of gamma rays is observed in a $5^{\circ}$ region around the Sun. We therefore place the first preliminary upper limits on the flux. In the future, with more data and improvements in HAWC's gamma-hadron separation, we will be able to make these limits stronger. Gamma rays from the Sun below 10 GeV have already been measured to be an order of magnitude higher in flux than the prediction. The enhancement has been attributed to magnetic fields and has been tested at higher energies with HAWC. Moreover, the hadronic cosmic rays producing TeV photons would also produce neutrinos, which in addition to constraining the mass composition of the cosmic ray spectrum also offers an opportunity to correlate measurements across telescopes. Finally, annihilation of WIMPs within the Sun has also been proposed as a potential source of gamma rays. While the preliminary limits are not strong enough to test the aforementioned mechanisms accurately, increased amount of data from HAWC at TeV energies will be useful in setting constraints on the underlying physics.
\section*{\centerline{Acknowledgements}}
\small
We acknowledge the support from: the National	Science	Foundation	(NSF);	the	
US	Department	of	Energy	Office	of	High-Energy	Physics;	the	Laboratory	Directed	
Research	and	Development	(LDRD)	program	of	Los	Alamos	National	Laboratory;	
Consejo	Nacional	de	Ciencia	y	Tecnolog\'{\i}a	(CONACyT),	M{\'e}xico	(grants	
271051,	232656,	260378,	179588,	239762,	254964,	271737,	258865,	243290,	
132197),	Laboratorio	Nacional	HAWC	de	rayos	gamma;	L'OREAL	Fellowship	for	
Women	in	Science	2014;	Red	HAWC,	M{\'e}xico;	DGAPA-UNAM	(grants	RG100414,	
IN111315,	IN111716-3,	IA102715,	109916,	IA102917);	VIEP-BUAP;	PIFI	2012,	
2013,	PROFOCIE	2014,	2015; the	University	of	Wisconsin	Alumni	Research	
Foundation;	the	Institute	of	Geophysics,	Planetary	Physics,	and	Signatures	at	Los	
Alamos	National	Laboratory;	Polish	Science	Centre	grant	DEC-2014/13/B/ST9/945;	
Coordinaci{\'o}n	de	la	Investigaci{\'o}n	Cient\'{\i}fica	de	la	Universidad	
Michoacana. Thanks	to	Luciano	D\'{\i}az	and	Eduardo	Murrieta	for	technical	
support.
\normalsize

\bibliography{bib}{}
\bibliographystyle{ieeetr}

\end{document}